\documentclass[11pt]{article}
\usepackage{makeidx}
\usepackage{amssymb}
\usepackage{stmaryrd}
\usepackage{graphicx}
\usepackage{amsmath}
\usepackage{palatino}
\usepackage{latexsym}
\newenvironment{proof}{\par \noindent {\bf Proof: }}{\begin{flushright}
$\Box$\end{flushright}\par \noindent}

\begin{document}
\title{On Computation of Error Locations and Values in Hermitian Codes}
\author{Rachit Agarwal\thanks{Rachit Agarwal (ragarwal@cs.ucc.ie) is with Centre for Efficiency-Oriented Languages (CEOL), Department of Computer Science, University College Cork, Ireland.}}
\date{(Draft Version, Submitted to ITW 2008) \\November 20, 2007}
\maketitle

\begin{abstract}
We obtain a technique to reduce the computational complexity associated with decoding of Hermitian codes. In particular, we propose a method to compute the error locations and values using an uni-variate error locator and an uni-variate error evaluator polynomial. To achieve this, we introduce the notion of Semi-Erasure Decoding of Hermitian codes and prove that decoding of Hermitian codes can always be performed using semi-erasure decoding. The central results are:
\begin{itemize}
\item [$\star$] Searching for error locations require evaluating an univariate error locator polynomial over $q^2$ points as in Chien search for Reed-Solomon codes.
\item [$\star$] Forney's formula for error value computation in Reed-Solomon codes can directly be applied to compute the error values in Hermitian codes.
\end{itemize}
The approach develops from the idea that transmitting a modified form of the information may be more efficient that the information itself. 
\end{abstract}
%
%
\section{Introduction}
The decoding problem in Hermitian codes is split into three main parts, \emph{viz.}, syndrome computation, computing the error locator polynomial and computation of locations and error values. The main persisting problem with Hermitian codes is the timing and computational complexity associated with the decoding of the codes. Efficient computation of error locator polynomial and error evaluator polynomial have been proposed by Koetter and O'Sullivan \cite{koetterd}-\cite{mike2}. Specifically, the computation of error locator polynomials can now be performed very efficiently \cite{koetterd}\cite{mike3}\cite{jian}.

Besides computing the error evaluator polynomial, most of the research has been done on computing the error evaluator polynomial and the values of the errors using Forney type formula used in Reed-Solomon (RS) codes \cite{koetter2}-\cite{mike4}. In terms of finding the locations of the errors, it has been widely believed that given the error locator polynomial, significant improvements can not be achieved in terms of finding the error locations. This is due to the fact that the only approach is to find the zeros of the error locator polynomial by evaluating it at all possible points on the curve. This search process has a high timing and computational complexity. Further, techniques to compute the error values involve either high order hasse derivatives over bivariate polynomials and evaluating those polynomials at every error value or to find the error locator which has a zero of multiplicity one at the error location \cite{errevk}. This is different to evaluating error values in Reed-Solomon (RS) codes where we get an univariate error locator polynomial.

In this paper, we present some further improvements for the computation of error values from the error evaluator polynomial. The proposed technique also gives significant reduction in complexity of computing the error locations given the error locator polynomial. In particular, we obtain a method to employ univariate polynomials which have as roots the locations of the errors and univariate error locator polynomial which allows to use exactly the Forney's formula for finding the values of the errors. 

To achieve this, we introduce the notion of semi-erasures, which is essentially when the decoder has a partial knowledge about the error locations in the received word. We give a method to identify partial knowledge about the error locations and show its applicability for any random error pattern. 
The proposed approach is based on an interesting observation that transmitting a modified form of the information may result in much lower decoding complexity than the actual information. 

The rest of the paper is organized as follows: Section \ref{hcasc} gives the basic definitions and code construction. We derive computability of an auxiliary codeword which we propose to transmit rather than the actual Hermitian codeword in Section \ref{cwtw}. We introduce the notion of semi-erasure decoding in Section \ref{statem} and prove its solvability. We show the existence of an univariate error locator polynomial for finding the error locations in Section \ref{locerr} and existence of an univariate error evaluator polynomial in Section \ref{errev}. Before closing the paper, we show that for our construction Forney formula from RS codes can directly be used for Hermitian codes as well in Section \ref{locerr}.
%
\section{Code Construction}
\label{hcasc} 
We consider codes from a Hermitian curve
\begin{displaymath}
\chi : x^{q+1} = y^q + y
\end{displaymath}
over a finite field $\mathbb{F}_{q^2}$. The space $L(mP_\infty)$ consists of all functions on $\chi$ that have a pole of multiplicity at most $m$ only at the unique point at infinity. We present the following proposition from \cite{notehs}\cite{generalrs2} without proof:\\
\\
{\bf{Proposition 1:}}\\
{\emph{The following set is a basis of $L(mP_\infty)$ for each $m \geq 0$}}
\begin{equation}
\label{basis}
\langle{x^ay^b : aq + b(q+1) \leq m,\ 0 \leq a,\ 0 \leq b < q}\rangle
\end{equation}

Let $y_0$ be an element of $\mathbb{F}_{q^2}$ that satisfies $y_0 + {y_0}^q = 1$. Then, according to \cite{generalrs2}, the affine rational points on $\chi$ can be represented as $(\eta, \eta^{q+1}y_0 + \beta_j)$, where $\eta \in \mathbb{F}_{q^2}$ and $\beta_j, j = 0, 1, \dots, q-1$ are the $q$ solutions in $\mathbb{F}_{q^2}$ for $y^q + y = 0$. Such a representation has been used in \cite{jian} recently to develop an efficient decoding algorithm for Hermitian codes. 

A similar, however, a little modified representation of the affine rational points has also been used in \cite{isithe} to develop an efficient encoding algorithm for Hermitian codes. The latter, shown in (\ref{affinep}), will play a major role in this paper: 
\begin{equation}
\label{affinep}
P_{\alpha, \beta} = (\alpha, \alpha^{q+1}(y_0 + \beta) + \delta(\alpha)\beta),
\end{equation}
where $\alpha$ and $\beta$ represent arbitrary elements in $\mathbb{F}_{q^2}$ and $\mathbb{F}_q$ respectively and $\delta$ is the Kronecker-delta function defined as 
\begin{displaymath}
\delta(\alpha) = \left\{ \begin{array}{ll}
0 & \textrm{if $\alpha \neq 0$}\\
1 & \textrm{if $\alpha = 0$}
\end{array} \right.
\end{displaymath}

Let $\epsilon$ be a primitive element in $\mathbb{F}_{q^2}$ and let $\gamma$ be a primitive element in $\mathbb{F}_q$. The positions in a codeword are labeled by the corresponding elements $\alpha$ = $\epsilon^i$, $\beta$ = $\gamma^j$ resulting in a codeword as a $q \times q^2$ matrix ${\bf{c}}$. Occasionally we will index elements in this array by elements of the fields $\mathbb{F}_{q^2}$ and $\mathbb{F}_{q}$, otherwise we index starting with $0$.

A Hermitian code $C(m)$ may be defined as
\begin{displaymath}
\{{\bf{c}} \in \mathbb{F}_{q^2}^{q^3}: \sum_{\alpha \in \mathbb{F}_{q^2}} \sum_{\beta \in \mathbb{F_q}} {\bf{c}}_{\beta,\alpha} f(P_{\alpha,\beta}) = 0,\ \forall f \in L(mP_\infty)\}
\end{displaymath}
For an in depth treatment of Hermitian codes, we refer to \cite{notehs}\cite{goppa}\cite{structure}.
%
\section{Codeword and Transmitted Word}
\label{cwtw}
The central idea of our approach is that transmitting a modified form of the information may be more efficient (in terms of decoding complexity and error correcting capability) than the information itself. This approach, though not very trivial, is fairly easy to formulate. In this section, we obtain such a modified information for Hermitian codeword. Let us call this auxiliary codeword. 

Let a received word ${\tilde{\textbf{y}}} = {\tilde{\textbf{r}}} + {\tilde{\textbf{e}}}$ be given such that all of ${\tilde{\textbf{y}}}$ (auxiliary received word), ${\tilde{\textbf{r}}}$ (auxiliary codeword) and ${\tilde{\textbf{e}}}$ are matrix of dimension $q \times q^2$, and where ${\tilde{\textbf{e}}}$ is an error vector with Hamming weight $t < d^*(C(m))/2$. We define a ``Mapping matrix'' ${\textbf{M}}$ of dimension $q \times q$ such that
\begin{equation}
\label{eqn_dbl_x} 
\begin{array}{ll}
{\textbf{y}} & = {\textbf{M}} \times {\tilde{\textbf{y}}}= {\textbf{M}} \times {\tilde{\textbf{r}}} + {\textbf{M}} \times {\tilde{\textbf{e}}} = {\textbf{r}} + {\textbf{e}}\\
\end{array}.
\end{equation}
where, ${\textbf{r}} \in C(m)$. Specifically, given a $q \times q^2$ matrix ${\bf{\tilde{r}}}$ with columns ${\bf{\tilde{r}}}_j$, we define a $q \times q^2$ matrix ${\bf{r}}$ with columns ${\bf{r}}_j$ as
\begin{equation}
\label{genr}
{\bf{r}}_j = M_j \times \tilde{{\bf{r}}}_j
\end{equation}
where, $M_j$ is chosen from the following array $\overline{M}$ of matrices of type $M$ and $M'$:
\begin{displaymath}
\overline{M} = (M_0, M_1, \dots, M_{q^2-1}),\
M_i = \left\{ \begin{array}{ll}
M' & \textrm{i = $q^2$-1 }\\
M & \textrm{otherwise}
\end{array} \right.
\end{displaymath}
The problem now is to show the existence of such a mapping matrix and an auxiliary codeword. 
\\
\\
{\bf{Proposition 2 [Computability of Auxiliary Codeword]:}}\\
{\emph{There always exist Mapping Matrices {\bf{M}} and {\bf{M'}}. Given the mapping matrices, there always exists an auxiliary codeword for a given Hermitian codeword which satisfies (\ref{eqn_dbl_x}).}}\\
\begin{proof}
In \cite{isithe}, it is shown that given the definition of affine points (\ref{affinep}), there always exist matrices $M^{-1}$ and $M'^{-1}$ which are Vandermonde matrices. Hence, they will always have inverses. From \cite{isithe}, these inverses of the form (\ref{invA}) and (\ref{invAi}) are essentially our mapping matrices. During the encoding of Hermitian codes using algorithm developed in \cite{isithe}, an auxiliary matrix ${\bf{\tilde{r}}}$ is also constructed. This proves that such mapping matrices and auxiliary codeword always exist.
\end{proof}
\begin{equation}
\label{invA}
\mathbf{M} =
\left( \begin{array}{cccc}
1- (y_0 + 0{)^{q-1}} & (y_0 + 0{)^{q-2}} & \ldots & (y_0 + 0{)^0}\\
1- (y_0 + 1{)^{q-1}} & (y_0 + 1{)^{q-2}} & \ldots & (y_0 + 1{)^0}\\
1- (y_0 + \gamma{)^{q-1}} & (y_0 + \gamma{)^{q-2}} & \ldots & (y_0 + \gamma{)^0}\\
\vdots & \vdots & \ddots & \vdots\\
1- (y_0 + {\gamma^{q-2}}{)^{q-1}} & (y_0 + {\gamma^{q-2}}{)^{q-2}} & \ldots & (y_0 + {\gamma^{q-2}}{)^0}\\
\end{array} \right)
\end{equation}
\begin{equation}
\label{invAi}
\mathbf{M'} =
\left( \begin{array}{cccccc}
1 & 0 & 0 & \ldots & 0 & -1\\
0 & -(1{)^{q-2}} & -(1{)^{q-3}} & \ldots & -(1{)^1} & -1\\
0 & -(\gamma{)^{q-2}} & -(\gamma{)^{q-3}} & \ldots & -(\gamma{)^1} & -1\\
\vdots & \vdots & \vdots & \ddots & \vdots\\
0 & -({\gamma^{q-2}}{)^{q-2}} & -({\gamma^{q-2}}{)^{q-3}} & \ldots & -({\gamma^{q-2}}{)^1} & -1\\
\end{array} \right)
\end{equation}
\section{Semi-Erasure Decoding}
\label{statem}
Decoding by errors and by erasures has been studied widely in coding theory literature, both for RS and Hermitian codes. In this section, we introduce the idea of semi-erasure decoding for Hermitian codes and show how it helps reducing the decoding complexity of Hermitian codes, specifically computation of error location and error values given the error locator polynomial.\\
\\
{\bf{Definition 1 [Semi-Erasure Decoding]:}}\\
{\emph{Given a Hermitian codeword, a semi-erasure is defined as the condition when knowledge of only one of the coordinates is required to identify the location of an error.}}\\
\\
We call it semi-erasure because it is neither a complete erasure decoding nor a complete error only decoding. Notice that in our definition of affine points, if we know the value of either of $\alpha$ or $\beta$, only the knowledge of the other is required to completely define the location of the point. If during our decoding process, we could identify the location of either of $\alpha$ or $\beta$, we will call it a semi-erasure decoding. We give a rather trivial definition which will simplify the further discussion:
\\
\\
{\bf{Definition 2: [Column Error Pattern]}}\\
{\it{We define an error pattern $e_{i,j}$ as a column error pattern if
\begin{displaymath}
e_{\alpha_0,\beta = 0} \neq 0 \Rightarrow e_{\alpha_0, \beta} \neq 0 \qquad \forall \beta \in \{1, \dots, q-1\}
\end{displaymath}}}
The following proposition gives a necessary condition for the fact that decoding of Hermitian codes can always be performed using semi-erasure decoding.\\
\\
{\bf{Proposition 3 [Semi-Erasure Decoding Can Always be Forced]:}}\\
{\emph{Given any error pattern introduced by the channel, transmitting the auxiliary codeword ${\bf{\tilde{r}}}$ will force Semi-Erasure Decoding.}}\\
\begin{proof}
Assume we always transmit an auxiliary codeword ${\bf{\tilde{r}}}$, the computability of which is shown in Proposition 2. Given any random error pattern, every error in ${\bf{\tilde{r}}}$ when operated over (\ref{genr}) will be converted into a column error as defined in Definition 2. Hence, we know $\beta$ for all errors occurring ($\beta = 0, 1, \dots, q-1$). Since $\beta$ is known, only the knowledge of $\alpha$ is required to completely determine the error location, which is Semi-Erasure Decoding as defined in Definition 3.
\end{proof}
Given that semi-erasure decoding can always be forced, we need the following theorem, which presents the relation between error weights in $\bf{\tilde{r}}$ and $\bf{r}$. We only sketch the proof here. 
\\
\\
{\bf{Theorem 1 [Solvability of Semi-Erasure Decoding]:}}\\
{\emph{The error weights for the auxiliary received word and corresponding Hermitian received word are equal. In other words, given a $q \times q^2$ auxiliary received word $\bf{\tilde{y}}$ with $t_{uc}$ uncorrectable errors, the errors can be corrected by the corresponding Hermitian code given that:}}
\begin{equation}
\label{rstoh}
t_{uc} \leq d_H
\end{equation}
{\emph{where, $d_H$ is the decoding capability of the Hermitian code.}}
\begin{proof}
(Sketch) In semi-erasure decoding, only one of the coordinates has to be solved for. It suffices to prove that the value of the unknown coordinate $c_0, c_1, \dots, c_{k-1}$ can be found, given the error correcting capability is $k$. The intuition behind the theorem is fairly clear. Given that the error correcting capability of the Hermitian code is $k$, in normal decoding process, we solve equations for $2k$ variables, $\alpha_{i}^{s}$ and $\beta_{i}^{s}$, whereas in this case, there are only $k$ variables given that either all of $\alpha_{i}^{s}$ or $\beta_{i}^{s}$ are known. Given the error correcting capability of the code is $k$, we can always solve for $2k$ variables in semi-erasure decoding. Hence the claim.
\end{proof}

This in itself is an interesting result since this gives us a hint as to why some of the error patterns require lesser check symbols for decoding of Hermitian codes. This is one of the effects noticed by O'Sullivan \cite{dbmdb2}.
\section{Error Location Computation}
\label{locerr}
Once an error locator polynomial has been computed in Koetter's decoding algorithm, the only known way to compute the error locations is through Chien Search, which implies computing the value of the error locator polynomial at all possible points on the curve. The error locations are then the zeros of the error locator polynomial.

This method of computing the error locations require evaluating a bivariate error locator polynomial at $q^3$ points of the curve. Even for moderate number of errors, the computational complexity of this operation is very high. In this section, we show how to reduce this complexity to that of Reed-Solomon codes (computing the polynomials at $q^2$ locations).\\
\\
{\bf{Theorem 2: [Uni-variate Error Locator Polynomial]\\}} {\emph{For semi-erasure decoding, the zeros of $\sigma(x,y)$ are equivalent to the zeros of an uni-variate polynomial $\psi(x)$. In general, $\psi(x)$ is of the form: 
\begin{equation}
\label{psix}
\psi(x) = f(x) \prod_{k : e_k \neq 0} (x - x_k)
\end{equation}
where $f(x) \neq 0$ at any error location.}}\\
\begin{proof}
Recall that for semi-erasure decoding, one of the coordinates defining the error positions will be known. For our construction, we know from proof of Proposition 3 that the values of $\beta$ for all errors are known. Given any bivariate error locator polynomial $\sigma(x, y)$ (from Koetter's decoding algorithm, say), we know that zeros of $\sigma(x, y)$ are equivalent to roots of $\sigma(x, x^{q+1}(y_0 + \beta) + \delta(x)\beta)$ using the definition of our affine points, where $\beta$ is known. This is an univariate polynomial in $x$. 

The zeros of this polynomial are the x-coordinate of the error positions, say $x_1, x_2, \dots, x_t$. The following statements can be made for $\psi(x)$:
\begin{itemize}
\item $\psi(x)$ must have a term $\prod_{k : e_k \neq 0} (x - x_k)$, for $x_1, x_2, \dots, x_t$ are the solutions.
\item $\psi(x)$ may have a term $f(x)$ which can not be zero at any error location.
\item $\psi(x)$ can not have a term of the form $(x - x_r)^s$ for any $x$ and any $s \geq 2$.
\end{itemize}
The first two statements are trivial. We only show the third one. Assume that $\psi(x)$ has a term of the form $(x - x_r)^s$ for some $s \geq 2$. This will have multiple solutions at $x = x_r$. Given such a $\psi(x)$, there will always exist a $\psi'(x)$ which has a solution at $x = x_r$ and will be of the form $\psi'(x) = \psi(x)/(x - x_r)^{s-1}$ and hence, lower pole order than $\psi(x)$. Since decoding algorithms compute an error locator with the lowest possible pole order, such a $\psi'(x)$ is not possible. Then, by contradiction, $\psi(x)$ can not have a term of the form $(x - x_r)^s$ for some $s \geq 2$.
\end{proof}

Having discussed the uni-variate nature of our error locator polynomial, the following theorem is a direct result of the fact that the x-coordinate in our affine points satisfy $x \in \mathbb{F}(q^2)$:\\
\\
{\bf{Proposition 4: [Reduced Complexity Error Location Computation]\\}} {\emph{The Chien search for Semi-erasure decoding requires evaluating the $\psi(x)$ polynomial only at $q^2$ points.}}\\
\\
Given the complexity of error locator polynomial in error-only decoding of Hermitian codes (pole order as high as $t + 2g + q - 1$), this is a significant reduction in computational steps for finding the error locations. This is due to the fact that we now need to evaluate the polynomial only at $q^2$ points (rather than $q^3$ for error-only decoding) and also, we are working with only one variable for evaluating the polynomial.

\section{Error Evaluation}
\label{errev}
The decoding complexity of Hermitian codes can be further reduced by using semi-erasure decoding. This improvement can be done in evaluating the values of the errors. In particular, semi-erasure decoding allows us to use Forney's formula directly for computing error values in Hermitian codes.\\
\\
{\bf{Lemma 1:}} {\emph{Given an error locator polynomial $\psi(x)$ of the form as in Theorem 2, the derivative $\psi'(x)$ with respect to $x$ can not be zero at any of the error locations.}}\\
\begin{proof}
If $\psi(x)$ is of the form as shown in (\ref{psix}), its derivative with respect to $x$ can be written as:
\begin{displaymath}
\psi'(x) = (f(x) \sum_{k : e_k \neq 0} \prod_{j \neq k} (x - x_j)) + f'(x) \prod_{k : e_k \neq 0} (x - x_k)
\end{displaymath}
For any error location $x_k$, 
\begin{equation}
\label{psip}
\psi'(x_k) = f(x_k) \sum_{k : e_k \neq 0} \prod_{j \neq k} (x_k - x_j)
\end{equation}
which is non-zero.
\end{proof}

Given the syndromes and the error locator polynomial $\sigma(x,y)$, it is known how to compute the error evaluator polynomial $\omega(x, y)$ \cite{errev}\cite{errevk}\cite{mike1}\cite{mike4}. Before deriving the applicability of Forney's formula in error value computation in Hermitian codes, we will need a simple Lemma.\\
\\
{\bf{Theorem 3: [Uni-variate Error Evaluator Polynomial]}}\\
{\emph{For semi-erasure decoding, an error evaluator polynomial $\omega(x, y)$ corresponding to $\sigma(x, y)$ is equivalent to an uni-variate polynomial $\Omega(x)$ corresponding to the uni-variate error locator polynomial $\psi(x)$, i.e., satisfies:
\begin{equation}
\Omega(x) = \psi(x)\ (S_e\mid_{\beta = \beta_j})
\end{equation}
Further, $\Omega(x)$ is of the form:
\begin{equation}
\label{omx}
\Omega(x) = f(x)\sum_{k : e_k \neq 0} e_k\ g(x, x_k)\  \prod_{j \neq k} (x - x_j)
\end{equation}
where $f(x) \neq 0$ at any error location and $g(x, x_k)\mid_{x = x_k} = 1$ at all error locations.}}\\
\begin{proof}
To prove this, we use an alternative definition of syndrome polynomial for an error vector $e$:
\begin{equation}
\label{se}
S_e = \sum_{k=1}^{n} e_kh_k
\end{equation}
where 
\begin{equation}
\label{hk}
h_k = \frac{1}{x-x_k}\frac{y^q + y - y_k^q - y_k}{y - y_k}
\end{equation}
We say that polynomials $\sigma, \omega \in \mathbb{F}_{q^2}[x,y]$ satisfy the key equation if:
\begin{displaymath}
\omega = \sigma S_e
\end{displaymath}
Since this is true for all points, without loss of generality we can say that for any $y = x^{q+1}(y_0 + \beta) + \delta(x)\beta$:
\begin{displaymath}
\omega\mid_y = (\sigma\mid_y) (S_e\mid_y)
\end{displaymath}
Since $\beta$ is known for semi-erasure decoding, we can reduce it to:
\begin{displaymath}
\omega\mid_{\beta} = (\sigma\mid_{\beta}) (S_e\mid_{\beta})
\end{displaymath}
which is equivalent to:
\begin{equation}
\label{univev}
\Omega(x) = \psi(x) S_e\mid_{\beta = \beta_j}
\end{equation}
To prove the second statement, we expand the fractions in (\ref{hk}):
\begin{equation}
h_k = \frac{1}{x-x_k}(y^{q-1} + 1 + y_ky^{q-2} + \dots + y_k^{q-2}y + y_k^{q-1})
\end{equation}
Notice that for any error position with y-coordinate $y_k$ and for fields of characteristic 2, it is easy to show that:
\begin{displaymath}
h_k = \frac{1}{x-x_k}
\end{displaymath}
which implies that substituting the explicit form of y-coordinate from (\ref{affinep}), $y = x^{q+1}(y_0 + \beta) + \delta(x)\beta$, we get:
\begin{displaymath}
h_k = \frac{1}{x-x_k} g(x, x_k)
\end{displaymath}
where $g(x) = 1$, when we substitute the value of x-coordinate. Plugging this in (\ref{se}), we get:
\begin{equation}
\label{fse}
S_e\mid_{\beta = \beta_j} = \sum_{k = 1}^n \frac{e_k\ g(x, x_k)}{x-x_k}
\end{equation}
Plugging $\psi(x)$ from (\ref{psix}) and $S_e\mid_{\beta = \beta_j}$ from (\ref{fse}) in (\ref{univev}), we get:
\begin{displaymath}
\Omega(x) = f(x)\prod_{k : e_k \neq 0} (x - x_k) \sum_{k = 1}^n \frac{e_k\ g(x, x_k)}{x-x_k}
\end{displaymath}
\begin{equation}
\Rightarrow \Omega(x) = f(x) \sum_{k : e_k \neq 0} e_k\ g(x, x_k)\  \prod_{j \neq k} (x - x_j)
\end{equation}
such that $f(x) \neq 0$ at any error location and $g(x, x_k)\mid_{x = x_k} = 1$ for all error locations.
\end{proof}

Finally we show that errors can be evaluated using a Reed-Solomon type error evaluation.\\
\\
{\bf{Theorem 4: [Forney's Formula for Hermitian Codes]}} {\emph{For any error location at point $P_k$ with x-coordinate $x_k$, error value can be calculated using the formula:
\begin{equation}
\label{errorf}
e_{x_k} = \dfrac{\Omega(x_k)}{\psi'(x_k)}
\end{equation}
where $\psi$ and $\Omega$ are uni-variate error locator and error evaluator polynomials and $\psi'(x)$ is derivative with respect to x.}}\\
\begin{proof}
Substituting the values for $\Omega(x)$ and $\psi'(x)$ from (\ref{omx}) and (\ref{psip}) respectively, the R.H.S. of the above equation (\ref{errorf}) becomes:
\begin{displaymath}
\frac{f(x) \sum_{k : e_k \neq 0} e_k\ g(x, x_k)\  \prod_{j \neq k} (x - x_j)}{f(x) \sum_{k : e_k \neq 0} \prod_{j \neq k} (x - x_j) + f'(x) \prod_{k : e_k \neq 0} (x - x_k)}
\end{displaymath}
This fraction at point $x = x_k$ and using constraint that $g(x, x_k)\mid_{x= x_k} = 1$, we get:
\begin{displaymath}
\frac{\Omega(x_k)}{\psi'(x_k)} = \frac{f(x_k) \sum_{k : e_k \neq 0} e_k \prod_{j \neq k} (x_k - x_j)}{f(x_k) \sum_{k : e_k \neq 0} \prod_{j \neq k} (x_k - x_j)} = e_k
\end{displaymath}
Hence the proof.
\end{proof}
\section{Discussion and Future Directions}
\label{close}
In this paper, we have shown that by carefully constructing the codes and transmitting a modified information codeword, we can significantly reduce the decoding complexity in Hermitian codes. Of particular interests are results concerning formation of univariate error locator and error evaluator polynomials to locate and evaluate the errors. It has been shown that the Chien search for Hermitian codes can be reduced to evaluating an univariate polynomial over $q^2$ points as opposed to a bivariate polynomial evaluation on $q^3$ points. Finally, it has been shown that Forney formula from Reed-Solomon codes can be used for evaluating the error values in Hermitian codes. Though it sounds a little not so obvious, but proof of Theorem 1 does indicate that using this construction, it might be possible to decode Hermitian codes using a single modified Berlekamp Massey algorithm. An answer to this might result in highly reduced complexity of decoding of Hermitian codes, thereby solving a long standing problem of enhancing the applicability of the strongest competitors of ubiquitous Reed-Solomon codes.
%

\end{document}